\documentclass[usenatbib]{mn2e}
\usepackage{graphicx}
\usepackage{mathptmx}
\def\AD{{\sc ad}}

\def\SNR(#1.#2)#3(#4.#5){{G#1${\cdot}$#2$#3$#4${\cdot}$#5}}
\title[Radio expansion and brightening of the SNR \SNR(1.9)+(0.3)]{The radio
expansion and brightening of the very young supernova remnant \SNR(1.9)+(0.3)}

\label{firstpage}

\author[Green et~al.]{D.~A.~Green,$^1$\thanks{email: {\tt dag@mrao.cam.ac.uk}}
                      S.~P.~Reynolds,$^2$ K.~J.~Borkowski,$^2$
                      U.~Hwang,$^3$ I.~Harrus,$^3$\newauthor and R.~Petre$^3$\\
 $^1$Astrophysics Group, Cavendish Laboratory, 19 J.~J.~Thomson Avenue,
 Cambridge CB3 0HE\\
 $^2$Department of Physics, North Carolina State University,
 Raleigh NC 27695-8202, U.S.A.\\
 $^3$NASA/GSFC, Code 660, Greenbelt, MD 2077, U.S.A.}

\date{Accepted ---; received ---; in original form 2008 April 10}

\pagerange{\pageref{firstpage}--\pageref{lastpage}}

\pubyear{2008}
\begin{document}

\maketitle

\begin{abstract}
Recent radio observations of the small Galactic supernova remnant
\SNR(1.9)+(0.3) made at 4.86~GHz with the VLA are presented, and compared with
earlier observations at 1.49~GHz which have a comparable resolution ($10 \times
4$~arcsec$^2$). These show that the radio emission from this remnant has
expanded significantly, by about 15~per~cent over 23 years, with a current
outer diameter of $\approx 92$~arcsec. This expansion confirms that
\SNR(1.9)+(0.3) is the youngest Galactic remnant yet identified, only about 150
years old at most. Recent, lower resolution, 1.43-GHz observations are also
discussed, and the integrated flux densities from these and the 4.86-GHz
observations are compared with earlier results. This shows that the integrated
flux density of \SNR(1.9)+(0.3) has been increasing recently.
\end{abstract}


\section{Introduction}

The small Galactic `shell' type supernova remnant (SNR) \SNR(1.9)+(0.3) was
identified by \citet{1984Natur.312..527G}, from Very Large Array (VLA)
observations of a sample of Galactic radio sources. It was revealed as a small
limb-brightened shell, brighter in the north, only just over 1~arcmin in
diameter, and is the smallest catalogued Galactic SNR
(\citealt{2004BASI...32..335G}\footnote{See also: {\tt
http://www.mrao.cam.ac.uk/surveys/snrs/} for an updated version (from 2006
April).}). Although its distance was not known at the time of its
identification, it was clear that it must be a young SNR, due to its small
physical size (e.g.\ even at distance as large as 17~kpc -- i.e.\ twice the
distance to the Galactic Centre -- its physical diameter would be $\approx
6$~pc, which is comparable to that of the remnant of Tycho's SN of
\textsc{ad}1572). As such \SNR(1.9)+(0.3) is one of few `young but distant'
SNRs, which are missing from current catalogues, due to selection effects
(e.g.\ \citealt{2005MmSAI..76..534G}), but perhaps also indicating a true
deficit \citep{2006A+A...450.1037T}. Searches for these missing young remnants
have had limited success (see \citealt{1984Natur.312..527G,
1984AJ.....89..819H, 1985MNRAS.216..691G, 1989AJ.....98.1358G,
1992AJ....104..704S, 2002MNRAS.335..114M}, see also
\citealt{2004MNRAS.354..827S}).

There are few published observations of \SNR(1.9)+(0.3) in the literature. VLA
images at 4.9-GHz and 1.5-GHz (from observations made in 1985) are presented in
\citet{1984Natur.312..527G} and \citet{2004BASI...32..335G} respectively.
\citet{1994MNRAS.270..835G} presents MOST observations at 843~MHz, but these
barely resolve the remnant. \citet{2004AJ....128.1646N} and
\citet{2004ApJS..155..421Y} present VLA observations, at 330~MHz and 1.5~GHz
respectively, but of limited quality or resolution. \SNR(1.9)+(0.3) has also
been detected in X-rays, with ASCA as AX~J1748.7$-$2709
(\citealt{2002ApJS..138...19S}).

\begin{table}
\caption{VLA observations of \SNR(1.9)+(0.3).\label{tab:obs}}
\begin{tabular}{lcc}\hline
                            &  1985 observations & 2008 observations \\ \hline
 date                       &    1985 April 16   &   2008 March 12   \\
 VLA array                  &         B          &        C          \\
 frequency                  &      1.49 GHz      &     4.86 GHz      \\
 time on source             &       25 min       &      29 min       \\
 primary calibrator         &       3C286        &      3C286        \\
 \quad assumed flux density &      14.70~Jy      &     7.49~Jy       \\
 secondary calibrator       &     B1829$-$106    &   J1751$-$251     \\
 \quad derived flux density &      0.927~Jy      &     0.569~Jy      \\ \hline
\end{tabular}
\end{table}

\begin{figure*}
%
%
\centerline{\includegraphics[width=\columnwidth]{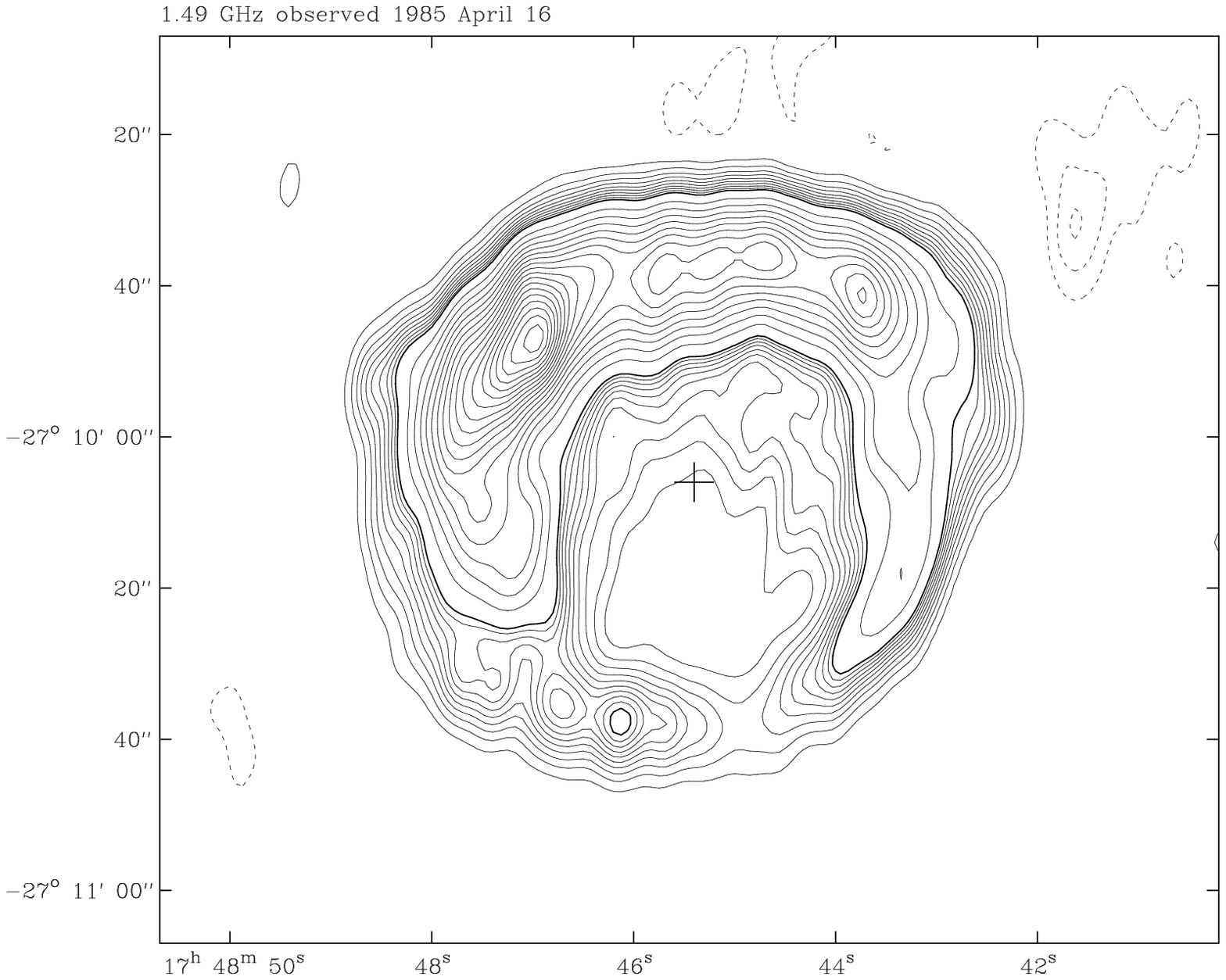} \quad
            \includegraphics[width=\columnwidth]{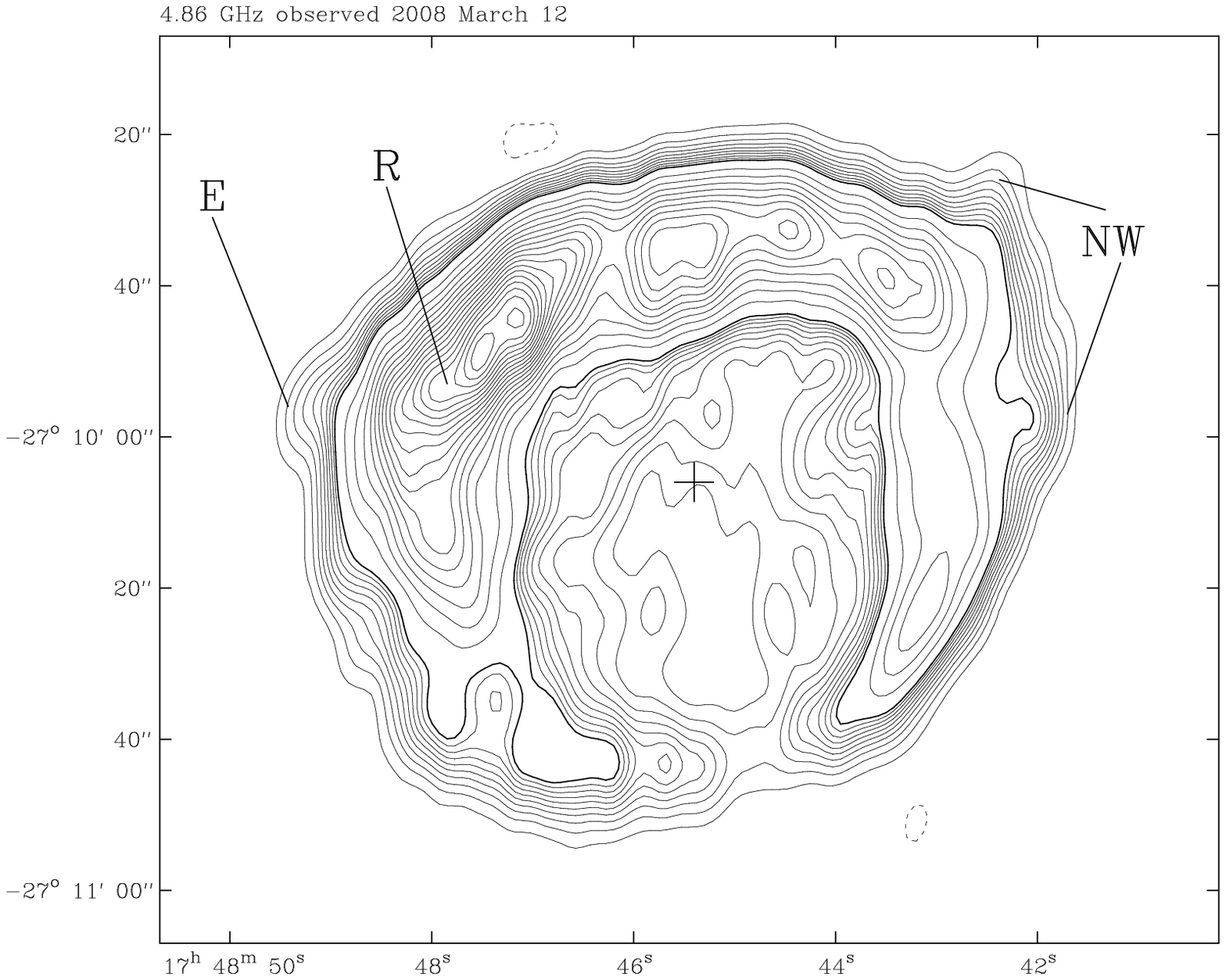}}
\caption{VLA observations of \SNR(1.9)+(0.3): (left) at 1.46~GHz from 1985 with
contour levels at $(-3, -2, -1, 1, 2, 3\dots 10, 15, 20\dots 95) \times
0.30$~mJy~beam$^{-1}$, and (right) at 4.86~GHz from 2008 with contour levels at
$(-3, -2, -1, 1, 2, 3\dots 10, 15, 20\dots 95) \times 0.17$~mJy~beam$^{-1}$.
The negative contours are dashed. Both images have a resolution of $10 \times
4$ arcsec$^2$ at $5.5^\circ$ W from N. The central crosses mark the centre
position used for radial profiles (see text and
Fig.~\ref{fig:radial}).\label{fig:match}}
\end{figure*}

Recently Chandra observations (\citealt{2008arXiv0803.1487R}, hereafter
Paper~I) have revealed that \SNR(1.9)+(0.3) is one of the few shell SNRs that
have line-free X-ray emission, dominated by synchrotron rather than thermal
emission. The size of the X-ray emission observed by Chandra was, unusually,
significantly larger than the then published radio observations, which were
interpreted in Paper~I as requiring a large expansion rate -- $\approx
16$~per~cent over 23 years -- which implies that \SNR(1.9)+(0.3) is a very
young SNR. Here we present recent new radio observations made with the VLA,
which in comparison with previous observations at a similar resolution, but a
different frequency, confirm the large expansion rate, and hence youth of this
remnant. We also discuss the temporal evolution of the flux density of
\SNR(1.9)+(0.3), and conclude that it has been brightening over recent decades.

\section{Observations}

\SNR(1.9)+(0.3) was recently observed with the VLA at 4.86~GHz in the C-array
configuration. The resolution of these observations closely match those of VLA
observations made in 1985 at 1.49~GHz in the B-array configuration (results
from which are shown in \citealt{2004BASI...32..335G}). Details of these
observations are given in Table~\ref{tab:obs}. The new observations were
processed using standard techniques in (classic) \textsc{aips} (see
\citealt{Bri94}). Obviously corrupted data were flagged, the flux scale
calibration was based on observations of 3C286, and antenna-based amplitude and
phase calibrations from observations of calibrator source(s) made nearby in
time were applied. The earlier 1.49-GHz observations were also reprocessed.
Although the time spent observing \SNR(1.9)+(0.3) was similar in each
observation, the details of the $uv$-coverage for the two observations are
different, due to differences in LST coverage, missing antennas, and flagging
of bad data. Consequently the synthesised beams are slightly different. For
expansion studies comparison images were made at a matched resolution, see
Fig.~\ref{fig:match}. These images show large expansion of the radio emission
from \SNR(1.9)+(0.3), which is discussed further in Section~\ref{sec:expand}.
The positional alignment of these images was checked from the apparent position
of a compact source that is about 1.4 arcmin north of the centre of the
remnant. The positions of this source were found to be in agreement to better
than 0.4~arcsec, which is small compared with the large changes seen between
the 1985 and 2008 image of \SNR(1.9)+(0.3). It should be noted, however, the
earlier 1.49-GHz observations probably do not have sufficient short baselines
to be able to fully image the largest scale emission from \SNR(1.9)+(0.3). The
VLA `Observational Status Summary'\footnote{Available via {\tt
http://www.vla.nrao.edu/}} lists the largest angular scale that `can be imaged
reasonably well' in a full synthesis as 2~arcmin for 1.49~GHz observations in
the B-array, and notes that this value should be halved for single snapshot
observations. Thus, given the angular size of \SNR(1.9)+(0.3) is over 1~arcmin,
the largest scale structure may not be fully imaged in the 1.49-GHz
observations. For the 4.86-GHz C-array observations the largest angular scale
well imaged with a full synthesis is 5~arcmin -- since the C-array is not an
exactly scaled version of B-array, but has better small baseline coverage -- so
even halving this for snapshot observations, the full emission from
\SNR(1.9)+(0.3) should be well imaged. Indeed, from Fig.\ref{fig:match}, the
1.49-GHz shows extended negative artefacts, which are features of the limited
small baseline coverage. The 4.86-GHz image is also better than the 1.49-GHz
image, in terms of its sensitivity: the nominal noise on these images, from the
r.m.s.\ away from bright emission, are about 0.038 and 0.012 mJy~beam$^{-1}$
at 1.49 and 4.86~GHz respectively.

Given the problem with missing large-scale emission in the 1.49~GHz
observations, we also made some short 1.43-GHz observations of \SNR(1.9)+(0.3)
in the same observation run as the 4.86-GHz observations, in order to
investigate temporal variations in the flux density of \SNR(1.9)+(0.3). These
consisted of two short ($\approx 4$~min) scans of \SNR(1.9)+(0.3), together
with adjacent calibration observations of 3C286 to set the overall flux density
scale (with an assumed flux density of 14.7~Jy), and of the nearby secondary
calibrator J1751$-$253 (for which a flux density of 1.18~Jy was derived). The
results from these observations are discussed below in
Section~\ref{sec:brighten}.

\begin{figure}
\centerline{\includegraphics[width=\columnwidth]{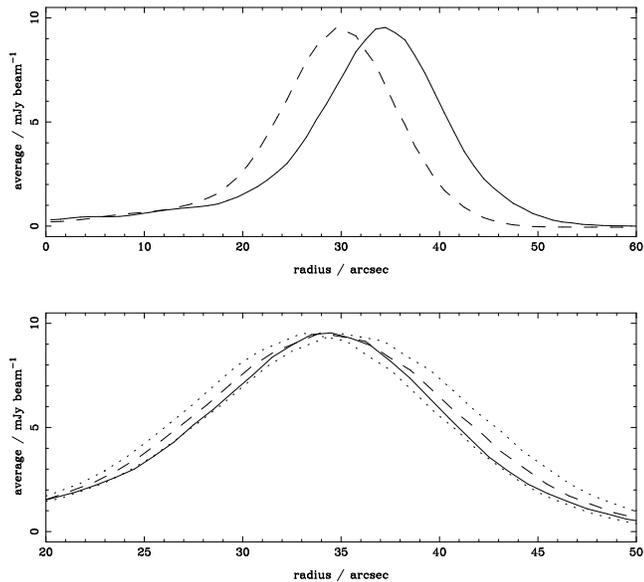}}
\caption{Top: radial profiles of radio emission from \SNR(1.9)+(0.3), from 1985
(dashed) and 2008 (solid). Bottom: scaled profile from 1985, expanded by 12, 15
and 18~per~cent (dotted, dashed, dotted), compared with the 2008 profile. (The
amplitudes of the profiles from 2008 have been multiplied by 1.65.)
\label{fig:radial}}
\end{figure}

\section{Discussion}

\subsection{Expansion and Structure}\label{sec:expand}

From Fig.~\ref{fig:match} it is clear that there is a large expansion of the
radio emission seen from \SNR(1.9)+(0.3) between 1985 and 2008. To quantify
this, radial profiles of the emission from each image were constructed -- see
Fig.~\ref{fig:radial} -- averaging over all angles, from a central position
which was adjusted to maximise the peak of the radial profile from the 2008
image. (This position is $17^{\rm h} 48^{\rm m} 45^{\rm s}.4$, $-27^\circ 10'
06''$, J2000.0, which close to the geometrical centre of the remnant.) A simple
re-scaling of the 1985 radial profile indicates an average expansion of
15~per~cent between 1985 and 2008, i.e.\ 0.65~per~cent yr$^{-1}$, is required
to match the positions of the peaks of the radial profiles
(Fig.~\ref{fig:radial}). This expansion rate is biased towards that appropriate
to the brighter N and NE portions of the remnant, since the radial profiles are
weighted by intensity.

This result confirms that a significant expansion is needed to match the extent
of the X-ray emission from the recent Chandra observations to older radio
observations, as found in Paper~I. Assuming free expansion, this implies an age
of 150 years for \SNR(1.9)+(0.3), which makes this the youngest known Galactic
SNR (but also see \citealt{2006ApJ...640L.179M} and
\citealt{2006ApJ...638..220R} for other possible very young SNRs which have
been reported). As the remnant will have undergone some deceleration, this age
is an upper limit. Very high X-ray absorption (Paper~I) suggests that
\SNR(1.9)+(0.3) is not much closer than the Galactic centre, while the very
high expansion would indicate a very large velocity if it were much further
away. Following Paper~I, we adopt a nominal distance of 8.5~kpc, which for an
age of $\la 150$~years and a outer radio diameter of 92~arcsec corresponds to a
mean expansion speed of $\ga 12,000$ km~s$^{-1}$.

\begin{figure}
%
%
%
\centerline{\includegraphics[width=\columnwidth]{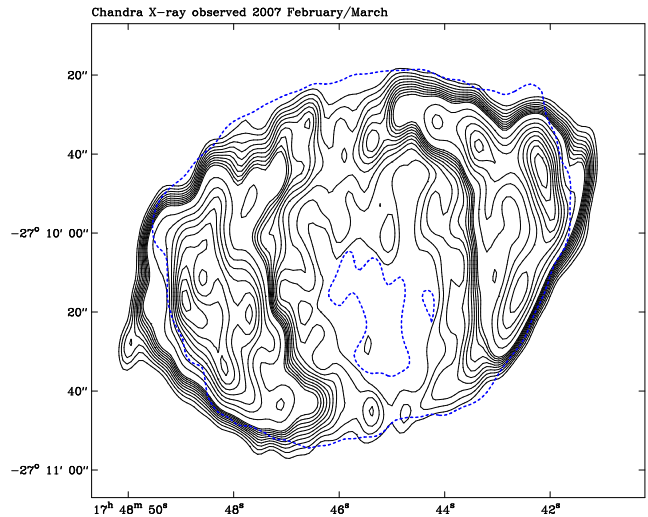}}
\caption{The black contours, at $(2.4, 3.0, 3.6\dots 7.8, 10.8, 13.8\dots 34.8)
\times 10^{-4}$ counts~s$^{-1}$beam$^{-1}$, are Chandra X-ray observations of
\SNR(1.9)+(0.3) (Paper~I), with the same resolution and scale as the radio
images shown in Fig.~\ref{fig:match}. (The background level away from the
remnant is $\approx 1.8 \times 10^{-4}$ counts~s$^{-1}$beam$^{-1}$.) The dotted
blue contour is the lowest contour of the 2008 4.86-GHz radio image shown in
Fig.~\ref{fig:match}.\label{fig:xray}}
\end{figure}

The interpretation of the radial profiles in simple terms of just a radial
scaling is, however, complicated by a variety of effects. First, recall that
the $uv$-coverage of these observations differ, particularly -- as noted
above -- that the 1.49-GHz observations may not well image the large scale
emission from \SNR(1.9)+(0.3). Second, the images are at different frequencies,
so that any change in spectral index across the remnant would appear as a
relative change in emission between the epochs (for example, a spectral index
change of 0.05 would correspond to a relative intensity change of 6~per~cent
between 1.49 and 4.86~GHz). Third, there are real changes in the emission over
the 23 years between the observations, so the changes seen are not just simple,
self-similar expansion. In the case of \SNR(1.9)+(0.3), we are not observing
the expansion of distinct features, such as the radio knots in Cas~A (see
\citealt{1995ApJ...441..307A}), or the well defined outer shock of Tycho's SNR
(e.g.\ \citealt{1985MNRAS.216..949T, 1997ApJ...491..816R}). From
Fig.~\ref{fig:match} there is evidence for changes in the shape and brightness
of the radio emission between 1985 and 2008. This is particularly noticeable in
the NW, where there is a clear extension outside the brighter rim of emission
from the remnant in the 2008 image (labelled `NW' in Fig.~\ref{fig:match}).
There is some indication of extension in the 1985 image, but at a much lower
level than in the 2008 image. The azimuth range of this outer extension
corresponds closely to where the X-ray emission shows a bright `ear' in this
region (Paper~I, see Fig.~\ref{fig:xray} above). In the E there is also an
faint extension outside the main emission particularly in the 2008 image
(labelled `E'), but over a much smaller azimuth range than the outer extension
in the NW. Also, the ridge line of the emission in the NE clearly differs
between the epochs, being concave (w.r.t.\ to the centre of the remnant) in the
1985 image, but slightly convex in the 2008 image (labelled `R'). If
extrapolated, this apparently points towards the faint E extension, which
corresponds to the northern edge of the X-ray emission outside the main rim in
the E. Also, the peaks of the bright ridge of the radio emission change between
the two epochs (e.g.\ the peak in the NE in the 1985 image splits into two
peaks in the 2008 image).

From Figs~\ref{fig:match} and \ref{fig:xray}, there are striking differences
between the overall structure of \SNR(1.9)+(0.3) at radio and X-ray
wavelengths, which may reflect differences in particle acceleration rates or
efficiencies at different locations. Although the outline of the remnant is
very nearly circular in the radio, the brightness of the radio emission varies
considerably in azimuth (with the bright NE rim being about fifty times
brighter than the weakest emission in the S). If a local density gradient were
invoked to explain the large N--S difference in radio brightness, then it is
not obvious why the remnant maintains such a circular outline. On the other
hand, the X-ray emission of the remnant shows a bipolar structure, brightest in
the E and W, which suggests the influence of a large scale magnetic field is
important for the efficiency of accelerating the particles responsible for the
X-ray emission, but \emph{not} for the those producing the radio emission. The
brightest emission in the E and particularly in the NW corresponds to
relatively faint radio emission. In the NW this is the outer extension seen in
the 2008 radio image, which presumably corresponds to the region of the outer
shock with the highest velocity.

\begin{table}
\tabcolsep 3.5pt
\caption{Radio flux densities for \SNR(1.9)+(0.3).\label{tab:fluxes}}
\begin{tabular}{rllclc}\hline
 $\nu$ &   $S$  & $\Delta S$ & observation & reference                   & notes \\
 / MHz &  / Jy  &   / Jy     & date(s)     &                             &       \\ \hline
  332  & 2.84   &    0.10    & 1986--1989  & \citet{2000AJ....119..207L} &       \\
  408  & 1.18   &    0.07    & 1969--1971  & \citet{1974AuJPh..27..713C} & a     \\
  843  & 1.0    &    0.05    & 1985--1991  & \citet{1994MNRAS.270..835G} & b     \\
  843  & 0.986  &    0.031   & 1997--2007? & \citet{2007MNRAS.382..382M} &       \\
 1400  & 0.748  &    0.038   & 1993--1996  & \citet{1998AJ....115.1693C} & c,d   \\
 1425  & 0.935  &    0.047   & 2008        & this work                   & c     \\
 2695  & 0.440  &    0.044   & 1981--1984? & \citet{1984A+AS...58..197R} &       \\
 4850  & 0.236  &    0.016   & 1990        & \citet{1994ApJS...90..179G} & e     \\
 4860  & 0.437  &    0.022   & 2008        & this work                   & c     \\
 4875  & 0.20   &    0.05    & 1974--1975  & \citet{1979A+AS...35...23A} & f     \\
 5000  & 0.20   &    0.03    & $\le$1975?  & Caswell et al.\ (\citeyear{1975AuJPh..28..633C}) & g \\ \hline
\end{tabular}
\begin{minipage}{\columnwidth}
\textbf{Notes:}\\
(a) The listed flux density error is based on the quoted 5.4~per~cent
statistical flux scale (for 1~Jy sources), plus 2~per~cent due to possible
systematic declination variations, added in quadrature.\\
(b) No flux density error is given, so a nominal value of 5~per~cent is used;
the observation dates are from \citet{Gra94}.\\
(c) These flux densities were obtained from integrations of images, from within
a polygon drawn around the remnant after removing a twisted plane fitted to
pixels lying around the edge of the polygon (see
\citealt{2007BASI...35...77G}). The integrated flux density was found to vary
by less than one per cent when a variety of polygons were used, but to be
cautious we assume a statistical error of 5~per~cent.\\
(d) The NVSS survey. \SNR(1.9)+(0.3) is also listed in the NVSS source
catalogue (Version 2.17): $58.8 \times 33.3$~arcsec$^2$ in extent, with $0.7453
\pm 0.0248$~Jy.\\
(e) The flux density is from the Vizier version (see
\citealt{2000A+AS..143...23O}) of the PMN Tropical zone catalogue allowing for
a general Gaussian width fit (see: \citealt{1993AJ....105.1666G}). The
published flux density from a fixed width fit is $0.175 \pm 0.014$~Jy.\\
(f) The flux densities of the sources in this survey are quoted to the nearest
0.1~Jy, so a flux density error of 0.05~Jy is used.\\
(g) The flux density error is from adding quoted errors of 10~per~cent for
uncorrected gain variations, 10~per~cent for positional errors (which would
lead to underestimates of the true value) and noise of 0.01~Jy, added in
quadrature.
\end{minipage}
\end{table}

\subsection{Brightening}\label{sec:brighten}

Given that \SNR(1.9)+(0.3) is such an apparently young SNR, then an obvious
question is how is its radio flux density evolving with time?
Table~\ref{tab:fluxes} lists the available flux densities for \SNR(1.9)+(0.3)
from the literature, and from the new observations presented here. (There is
also a flux density for \SNR(1.9)+(0.3) at 365~MHz from the Texas survey,
\citealt{1996AJ....111.1945D}, but this has not been listed as it is not
expected that it will represent the total flux of \SNR(1.9)+(0.3), which is
resolved by the instrument.) The available observations have been made with a
variety of instruments and resolutions, and may not be on consistent flux
density scales, making direct comparisons difficult. Nevertheless, there is
evidence that the flux density of \SNR(1.9)+(0.3) has been \emph{increasing}
over recent decades.
\begin{enumerate}
\item \citet{1994MNRAS.270..835G} noted that his 843-MHz flux density was
larger than expected from extrapolation between the earlier flux densities at
408~MHz and 5.0~GHz. \citeauthor{1994MNRAS.270..835G} suggested that this
discrepancy was due to the fact that the earlier 408-MHz and 5.0-GHz flux
densities were derived assuming the source was unresolved by the $\approx
3$~arcmin beams of both sets of observations, and so the observed peak flux
density would underestimate the integrated flux density. However, even if the
source were 1.2~arcmin in extent, as it was when identified by
\citet{1984Natur.312..527G}, this effect would be less than about 10~per~cent
for an $\approx 3$~arcmin beam, which is not sufficient to explain this
discrepancy. Also \citeauthor{1994MNRAS.270..835G} noted that recent -- i.e.\
presumably around 1994 -- 4.85-GHz observations (from `Haynes et~al.\ in
preparation', which we have not been able to identify in the literature) give a
flux density of 0.3~Jy for \SNR(1.9)+(0.3), which is above the earlier 5.0-GHz
flux density reported by \citet{1975AuJPh..28..633C}.
\item The 332-MHz flux density is considerably larger than the earlier value at
408~MHz (\citeauthor{1974AuJPh..27..713C}). (Note that
\citeauthor{2000AJ....119..207L} quote a 1.4-GHz flux density from NVSS with an
implausibly large error, i.e.\ $0.747\pm 0.250$~Jy, and hence also give very
large error in their derived 332~MHz to 1.4~GHz spectral index, i.e.\ $0.93 \pm
0.23$.)
\item The integrated flux density of \SNR(1.9)+(0.3) at 4.86~GHz from the
observations described above is significantly larger than any of the values
available in the literature at similar frequencies, even bearing in mind the
large errors in some of the earlier flux densities at $\approx 5$~GHz, and the
possibility of systematic differences in the flux density scales between the
different observations.
\item The 1.43-GHz flux density from our recent observations is about larger
than the previously available value at 1.4~GHz from 13~years earlier, by a
factor of $1.25 \pm 0.09$, which implies a flux density increase of about $\sim
2$~per~cent yr$^{-1}$.
\end{enumerate}
\noindent Combining the integrated flux densities from the contemporaneous, new
1.45 and 4.86-GHz observations gives a spectral index, $\alpha$, for the radio
emission from \SNR(1.9)+(0.3) (here defined in the sense that flux density
scales with frequency as $S \propto \nu^{-\alpha}$) of $0.62$. The error in
this spectral index is $\pm 0.06$ using the nominal 5~per~cent uncertainties in
the individual flux densities, not including any possible systematic
uncertainties in the relative flux density scales of the observations. However,
is it difficult to reconcile this spectral index with all of the other
observations in the literature. If \SNR(1.9)+(0.3) is indeed brightening, then
the spectral index between 332~MHz and 1.4~GHz would be larger than the value
of $0.93$ obtained by \citeauthor{2000AJ....119..207L}, as the 1.40-GHz was
obtained after the 332-MHz data. Also, the two 843-MHz flux densities do not
show obvious evidence for brightening. Excluding the 332-MHz flux density, it
is possible to obtain a reasonable fit of a spectrum with $\alpha \approx 0.7$
brightening at $\approx 2$~per~cent~yr$^{-1}$ to the available flux densities
for \SNR(1.9)+(0.3). The model fits reported in Paper~I assumed a flux density
at 1~GHz of 0.9 Jy. Increasing this to 1.2~Jy results in slightly steeper
values of the radio-to-X-ray spectral index. However, the new values of both
the spectral index and the rolloff frequency are still consistent with the
earlier values at the 90~per~cent confidence level.

The fact that \SNR(1.9)+(0.3) is brightening is not unexpected if it is indeed
only a hundred years or so old. Radio observations of supernovae show
brightening emission over time-scales of up to a year or so after the optical SN
due to decreasing opacity. Thereafter these radio supernovae (RSN) show steady
decline in emission (see, for example, \citealt{1986ApJ...301..790W,
2002ARA+A..40..387W}). Searches for radio emission from older RSN, up to about
80 years old (see \citealt{2002ApJ...573..306E}) mostly provide upper limits
for the radio emission, indicating that the luminosities of these older RSNe
are less than that of the brightest known Galactic SNR. Thus, in order to
evolve to match the observed properties of Galactic SNRs, it is expected that
radio emission from young SNRs will brighten, probably on time-scales of about a
century, when they have swept up sufficient interstellar medium (e.g.\
\citealt{1973MNRAS.161...47G, 1984MNRAS.207..745C}). (SN1987A is brightening at
radio wavelengths, as it interacts with ring of material very near the
explosion; \citealt{2002PASA...19..207M}.) For a current angular size of
$92$~arcsec, and a 1-GHz flux density of 1.2~Jy, the surface brightness of
\SNR(1.9)+(0.3) is $\approx 7.7 \times 10^{-20}$ W~m$^{-2}$Hz$^{-1}$sr$^{-1}$.
This means that \SNR(1.9)+(0.3) is brighter than the remnant of the SN of
{\AD}~1006, somewhat fainter than the remnants of Tycho's and Kepler's SNe of
{\AD}~1572 and 1604, but is considerably fainter than Cas~A, which is otherwise
the youngest (approximately 300 hundred years old) and brightest known Galactic
SNR (e.g.\ \citealt{2004BASI...32..335G}).

Predictions can be made for brightness increases based on simple assumptions
about particle acceleration and possible magnetic-field amplification. For
power-law behaviour of source radius with time, and of ejecta and ambient
density with radius, power law time dependences can be found for the
relativistic particle energy density (assuming a constant efficiency, i.e.\
$u_e \propto \rho v_s^2$), magnetic field (assuming either flux freezing
upstream or amplification with a different efficiency), and therefore the
synchrotron luminosity: $S_\nu \propto t^l$. Once substantial deceleration has
begun (i.e.\ once a reverse shock has formed, within a few years of the
explosion), almost all models decrease in luminosity with time. For constant
ambient density the drop is slowest; for constant ambient magnetic field as
well, the luminosity can actually rise, roughly as $t^{0.9}$ for the type~Ia SN
model of \citet{1982ApJ...258..790C}; all other cases drop with time. While
these are simplistic estimates, they do demonstrate that a {\sl growing} radio
flux density requires special conditions.

If G1.9+0.3 is now $\sim 100$ yr old, the observed flux increase by a factor of
1.25 over 13 years gives $l = 1.6$, far larger than the largest estimates
above. While other explanations cannot be ruled out, the most natural
explanation is that the efficiency with which shock energy goes into
relativistic electrons and/or magnetic field must be \emph{increasing with
time.}

\section{Conclusions}

Comparison of VLA observations made in 1985 and 2008 confirm that
\SNR(1.9)+(0.3) has a large expansion rate of $\approx
0.65$~per~cent~yr$^{-1}$. This means that this is a very young supernova
remnant, at most only 150 years old, and hence the youngest known Galactic
supernova remnant. There is also evidence that the this remnant has been
brightening over the last few decades at radio wavelengths, suggesting that the
efficiency of particle acceleration and/or magnetic-field amplification has
been increasing. Further high resolution, multi-epoch radio observations are
required to study \SNR(1.9)+(0.3)'s dynamics and structure in detail, and
monitor the temporal evolution of its flux density.

\section*{Acknowledgements}

The National Radio Astronomy Observatory is a facility of the National Science
Foundation operated under cooperative agreement by Associated Universities,
Inc. We are very grateful to the NRAO for the prompt scheduling of the
`Exploratory Time' proposal for the new VLA observations presented here. This
research has made use of NASA's Astrophysics Data System Bibliographic
Services.

\setlength{\labelwidth}{0pt} 

\bsp

\label{lastpage}
\end{document}